\documentstyle[fleqn,run2col,epsfig]{article}

\newcommand{\AmS}{{\protect\the\textfont2
  A\kern-.1667em\lower.5ex\hbox{M}\kern-.125emS}}

\title{Lepton Pair Production at the LHC and the Gluon Density in the Proton}
\author{E.~L.~Berger\address{High Energy Physics Division, Argonne National
        Laboratory, Argonne, IL 60439, USA}%
        \thanks{Supported by the U.S.\ Department of Energy,
        Division of High Energy Physics, under Contract
        W-31-109-ENG-38.}
        and 
        M.~Klasen\address{II.\ Institut f\"ur Theoretische
        Physik, Universit\"at Hamburg, Luruper Chaussee 149, D-22761
        Hamburg, Germany}%
        \thanks{Supported by Bundesministerium f\"ur Bildung
        und Forschung under Contract 05 HT9GUA 3, by Deutsche
        Forschungsgemeinschaft under Contract KL 1266/1-1, and by the
        European Commission under Contract ERBFMRXCT980194.}}
       
\begin{document}

\begin{abstract}
The hadroproduction of lepton pairs with mass $Q$ and finite transverse
momentum $Q_T$ is dominated by quark-gluon scattering in the region $Q_T > Q/2$. 
This feature provides a new independent method for constraining the gluon density 
with data at hadron collider energies.  Predictions are provided at the energy of 
the LHC.  
\end{abstract}

\maketitle

\vspace*{-7.2cm} \noindent ANL-HEP-CP-00-009 \\
\noindent DESY 00-032
\vspace*{ 5.7cm}

The production of lepton pairs in hadron collisions $h_1h_2\rightarrow\gamma^*
X;\gamma^*\rightarrow l\bar{l}$ proceeds through an intermediate
virtual photon via $q {\bar q} \rightarrow \gamma^*$, and the subsequent 
leptonic decay of the virtual photon. Interest in this Drell-Yan process is 
usually focussed on lepton pairs with large mass $Q$ which justifies the application 
of perturbative QCD and allows for the extraction of the antiquark density in 
hadrons \cite{Drell:1970wh}.  Prompt photon production 
$h_1h_2\rightarrow\gamma X$ can be calculated in perturbative QCD if the transverse 
momentum $Q_T$ of the photon is sufficiently large. Because the quark-gluon Compton 
subprocess is dominant, $g q \rightarrow \gamma X$, this reaction provides essential 
information on the gluon density in the proton at large $x$ \cite{Martin:1998sq}. 
Alternatively, the gluon density can be 
constrained from the production of jets with large transverse momentum at hadron 
colliders \cite{Lai:1999wy}.

In this report we exploit the fact that, along prompt photon production, lepton pair
production is dominated by quark-gluon scattering in the region $Q_T>Q/2$.
This realization means that new independent constraints on the gluon density 
may be derived from Drell-Yan data in kinematical regimes that are accessible 
at the Large Hadron Collider (LHC) but without the theoretical and experimental 
uncertainties present in the prompt photon case.

In leading order (LO) QCD, two partonic subprocesses contribute to the
production of virtual and real photons with non-zero transverse momentum:
$q\bar{q}\rightarrow\gamma^{(*)}g$ and $qg\rightarrow\gamma^{(*)}q$.
The cross section for lepton pair production is related to the cross section
for virtual photon production through the leptonic branching ratio of the
virtual photon $\alpha/(3\pi Q^2)$. The virtual photon cross section reduces
to the real photon cross section in the limit $Q^2\rightarrow 0$.

The next-to-leading order (NLO) QCD corrections arise from virtual one-loop
diagrams interfering with the LO diagrams and from real emission diagrams. At
this order $2 \rightarrow 3$ partonic processes with incident gluon pairs $(gg)$, 
quark pairs $(qq)$, and non-factorizable quark-antiquark $(q\bar{q}_2)$ processes 
contribute also.  
An important difference
between virtual and real photon production arises when a quark emits a
collinear photon. Whereas the collinear emission of a real photon leads to a
$1/\epsilon$ singularity that has to be factored into a fragmentation
function, the collinear emission of a virtual photon yields a finite
logarithmic contribution since it is regulated naturally by the photon
virtuality $Q$. In the limit $Q^2\rightarrow 0$ the NLO virtual photon
cross section reduces to the real photon cross section if this logarithm is
replaced by a $1/\epsilon$ pole. A more detailed discussion can be found
in Ref.~\cite{Berger:1998ev}.

The situation is completely analogous to hard
photoproduction where the photon participates in the scattering in the initial
state instead of the final state. For real photons, one encounters an
initial-state singularity that is factored into a photon structure function.
For virtual photons, this singularity is replaced by a logarithmic dependence
on the photon virtuality $Q$ \cite{Klasen:1998jm}.

A remark is in order concerning the interval in $Q_T$ in which our analysis is 
appropriate.  In general, in two-scale situations, a series of logarithmic 
contributions will arise with terms of the type $\alpha_s^n \ln^n (Q/Q_T)$.  Thus, 
if either $Q_T >> Q$ or $Q_T << Q$, resummations of this series must be considered. 
For practical reasons, such as event rate, we do not venture into the domain 
$Q_T >> Q$, and our fixed-order calculation should be adequate.  On the 
other hand, the cross section is large in the region $Q_T << Q$.  In previous 
papers~\cite{Berger:1998ev}, we compared our cross sections with available 
fixed-target and collider data on massive lepton-pair production, and we were able
to establish that fixed-order perturbative calculations, without resummation, 
should be reliable for $Q_T > Q/2$.  At smaller values of $Q_T$, non-perturbative 
and matching complications introduce some level of phenomenological ambiguity.  For 
the goal we have in mind, viz., contraints on the gluon density, it would appear 
best to restrict attention to the region $Q_T \geq Q/2$, but below $Q_T >> Q$.

We analyze the invariant cross section $Ed^3\sigma/dp^3$ averaged over the 
rapidity interval -1.0 $<y<$ 1.0.  We integrate the cross section over various 
intervals of pair-mass $Q$ and plot it as a function of the transverse 
momentum $Q_T$.  Our predictions are based on a NLO QCD 
calculation \cite{Arnold:1991yk} and are evaluated in the $\overline{\rm MS}$ 
renormalization scheme. The renormalization and factorization scales are set to 
$\mu=\mu_f= \sqrt{Q^2+Q_T^2}$. If not stated otherwise, we use the CTEQ4M
parton distributions \cite{Lai:1997mg} and the corresponding value of
$\Lambda$ in the two-loop expression of $\alpha_s$ with four flavors (five if
$\mu>m_b$). The Drell-Yan factor $\alpha/(3\pi Q^2)$ for the decay of the
virtual photon into a lepton pair is included in all numerical results.

\begin{figure}[htb]
 \begin{center}
  {\unitlength1cm
  \begin{picture}(7.6,10.5)
   \epsfig{file=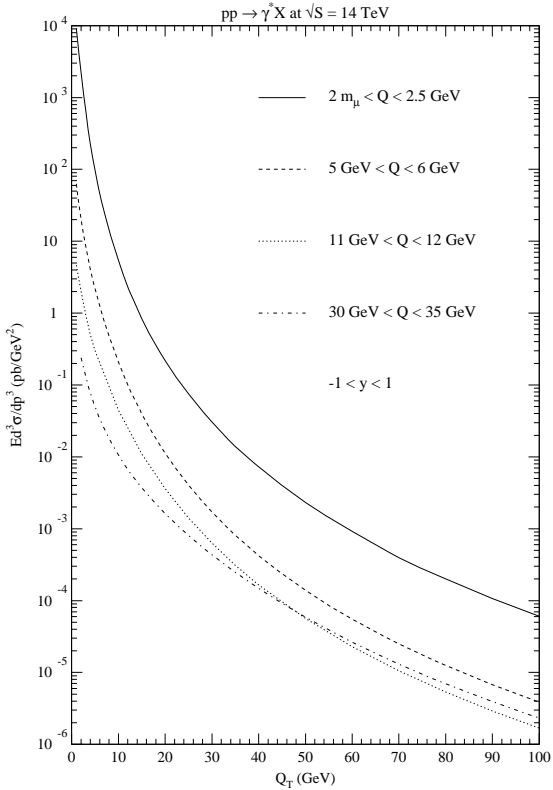,bbllx=60pt,bblly=100pt,bburx=495pt,bbury=725pt,%
           height=10.5cm}
  \end{picture}}
 \end{center}
\vspace*{-1cm}
\caption{Invariant cross section $Ed^3\sigma/dp^3$ as a function of $Q_T$
for $pp \rightarrow \gamma^* X$ at $\sqrt{S}=14$ TeV.}
\label{fig:1}
\end{figure}

In Fig.~\ref{fig:1} we display the NLO QCD cross section for lepton pair
production at the LHC at $\sqrt{S}=14$ TeV as a function of $Q_T$ for
four regions of $Q$ 
chosen to avoid resonances, {\it i.e.\ } from threshold to $2.5$ GeV,
between the $J/\psi$ and the $\Upsilon$ resonances, above the $\Upsilon$'s,
and a high mass region. The cross section falls both with the mass of the
lepton pair $Q$ and, more steeply, with its transverse momentum $Q_T$.
The initial LHC luminosity is expected to be 10$^{33}$ cm$^{-2}$ s$^{-1}$, or
10 fb$^{-1}$/year, and to reach the design luminosity of 10$^{34}$ cm$^{-2}$
s$^{-1}$ after three or four years. Therefore it should be possible to analyze
data for lepton pair production to at least $Q_T\simeq 100$ GeV where one
can probe the parton densities in the proton up to $x_T = 2Q_T/\sqrt{S}\simeq
0.014$. The UA1 collaboration measured the transverse momentum distribution
of lepton pairs at $\sqrt{S}=630$ GeV to $x_T=0.13$ \cite{Albajar:1988iq},
and their data agree well with our expectations \cite{Berger:1998ev}.

\begin{figure}[htb]
 \begin{center}
  {\unitlength1cm
  \begin{picture}(7.6,10.5)
   \epsfig{file=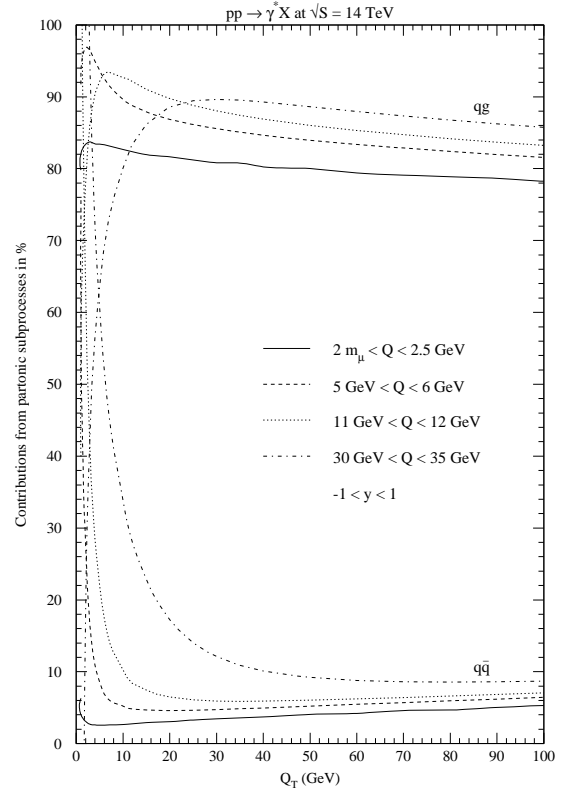,bbllx=60pt,bblly=100pt,bburx=495pt,bbury=725pt,%
           height=10.5cm}
  \end{picture}}
 \end{center}
\vspace*{-1cm}
\caption{Contributions from the partonic subprocesses $qg$ and $q\bar{q}$ to
the invariant cross section $Ed^3\sigma/dp^3$ as a function of $Q_T$
for $pp\rightarrow \gamma^* X$ at $\sqrt{S}$ = 14 TeV. The
$qg$ channel dominates in the region $Q_T > Q/2$.}
\label{fig:2}
\end{figure}

The fractional contributions from the $qg$ and $q\bar{q}$ subprocesses 
through NLO are shown in Fig.~\ref{fig:2}. It is evident 
that the $qg$ subprocess is the most important subprocess as long as
$Q_T > Q/2$. 
The dominance of the $qg$ subprocess increases somewhat with $Q$,
rising from over 80 \% for the lowest values of $Q$ to about 90 \%
at its maximum for $Q \simeq$ 30 GeV.
Subprocesses other than those initiated by the $q\bar{q}$ and
$q g$ initial channels are of negligible import.

\setcounter{footnote}{0}

The full uncertainty in the gluon density is not known.  We estimate the 
sensitivity of LHC experiments to the gluon density in the proton from
the variation of different recent parametrizations. We choose the latest
global fit by the CTEQ collaboration (5M) as our point of reference
\cite{Lai:1999wy} and compare results to those based on their preceding analysis 
(4M\cite{Lai:1997mg}) and on a fit with a higher gluon density (5HJ) intended to
describe the CDF and D0 jet data at large transverse momentum.  We also compare 
to results based on global fits by MRST \cite{Martin:1998sq}, who provide three 
different sets with a central, higher, and lower gluon density, and to GRV98
\cite{Gluck:1998xa}\footnote{In this set a purely perturbative generation of
heavy flavors (charm and bottom) is assumed. Since we are working in a massless
approach, we resort to the GRV92 parametrization for the charm contribution
\cite{Gluck:1992ng} and assume the bottom contribution to be negligible.}.

\begin{figure}[htb]
 \begin{center}
  {\unitlength1cm
  \begin{picture}(7.6,10.5)
   \epsfig{file=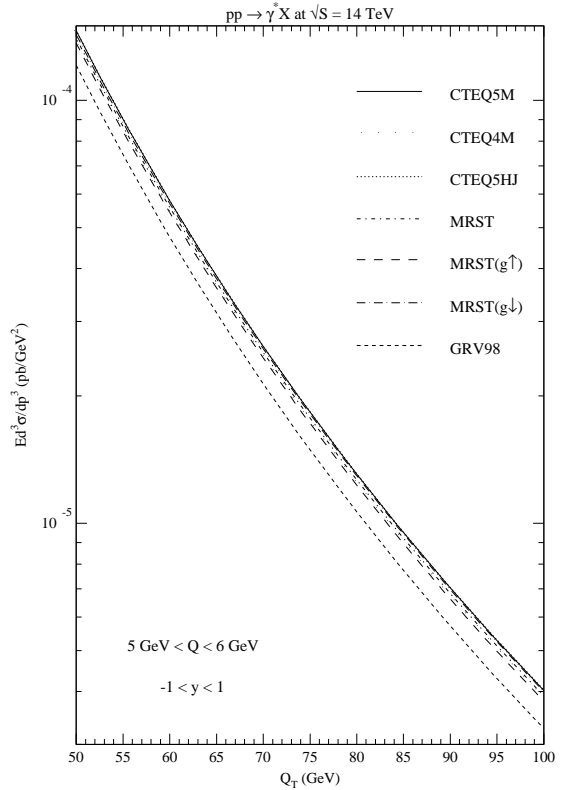,bbllx=60pt,bblly=100pt,bburx=495pt,bbury=725pt,%
           height=10.5cm}
  \end{picture}}
 \end{center}
\vspace*{-1cm}
\caption{Invariant cross section $Ed^3\sigma/dp^3$ as a function of $Q_T$
for $pp \rightarrow \gamma^* X$ at $\sqrt{S}=14$ TeV in the
region between the $J/\psi$ and $\Upsilon$ resonances. The largest differences
from CTEQ5M are obtained with GRV98 (minus 18 \%).}
\label{fig:3}
\end{figure}

In Fig.~\ref{fig:3} we plot the cross section for lepton pairs with mass 
between the
between $Q_T=50$ and 100 GeV ($x_T = 0.007 \dots 0.014$). For the CTEQ
parametrizations we find that the cross section increases from 4M to 5M by 5
\% and does not change from 5M to 5HJ in the whole $Q_T$-range. The largest
differences from CTEQ5M are obtained with GRV98 (minus 18 \%).

The theoretical uncertainty in the cross section can be estimated by varying
the renormalization and factorization scale $\mu=\mu_f$ about the central
value $\sqrt{Q^2+Q_T^2}$. 
In the region between the $J/\psi$ and $\Upsilon$ resonances,
the cross section drops from $\pm 39\%$ (LO) to
$\pm 16\%$ (NLO) when $\mu$ is varied over the 
interval interval $0.5 < \mu/\sqrt{Q^2+Q_T^2} < 2$.   
The $K$-factor ratio (NLO/LO) is approximately 1.3 at 
$\mu/\sqrt{Q^2+Q_T^2} = 1$.

We conclude that the hadroproduction of low mass lepton pairs is an advantageous
source of information on the parametrization and size of the gluon density.
With the design luminosity of the LHC, regions of $x_T \simeq 0.014$ should
be accessible. The theoretical uncertainty has been estimated from the scale
dependence of the cross sections and found to be small at NLO QCD.

\end{document}